\documentclass[aps,prb,twocolumn,reprint,
			   groupedaddress,superscriptaddress,
			   amsfonts,amssymb,amsmath,
			   citeautoscript,showpacs,
			   a4paper
			   ]{revtex4-1}

\usepackage{relsize}

\usepackage{microtype} 

\newsavebox{\oldell}
\savebox{\oldell}{\ensuremath{\ell}}

\usepackage{bm} 

\usepackage{xcolor}
\usepackage{graphicx}
\usepackage{tikz}
\renewcommand*{\ell}{\usebox{\oldell}}
\usepackage[shadow,textsize=small,textwidth=1.8cm,
backgroundcolor=green!20!white,
linecolor=black]{todonotes}
\usepackage{hyperref}
\hypersetup{
    colorlinks,
    linkcolor={blue!75!black!80!yellow},
    citecolor={blue!75!black!80!yellow},
    urlcolor={blue!75!black!80!yellow}
}

\usepackage[centering,hmargin=1.85cm,tmargin=31mm,bmargin=37mm]{geometry}

\graphicspath{{./graphicx/}}
\color{black!88!white}

\makeatletter
\renewcommand{\fnum@figure}{\figurename~\thefigure\ (color online)}
\newcommand{\bepsilona}{\mathlarger{\bm \varepsilon}}
\newcommand{\bepsilonb}{{\mathlarger{\mathlarger{\bm \varepsilon}}}}
\makeatother

\begin{document}
\title{\color{black!88!white}Quantum plasmonics: An energy perspective to quantify nonclassical effects}

\author{Wei~Yan}
\email[]{wyan@fotonik.dtu.dk}
\affiliation{Department of Photonics Engineering, Technical University of Denmark, DK-2800 Kgs. Lyngby, Denmark}
\affiliation{Center for Nanostructured Graphene, Technical University of Denmark, DK-2800 Kgs. Lyngby, Denmark}
\author{N.~Asger~Mortensen}
\email[]{asger@mailaps.org}
\affiliation{Department of Photonics Engineering, Technical University of Denmark, DK-2800 Kgs. Lyngby, Denmark}
\affiliation{Center for Nanostructured Graphene, Technical University of Denmark, DK-2800 Kgs. Lyngby, Denmark}

\date{\today}


\begin{abstract}\color{black!88!white}
Plasmons are commonly interpreted with classical electrodynamics, while nonclassical effects may influence the dynamics of plasmon resonances as the plasmon confinement is approaching the few--nanometer scale.
However, an unambiguous approach to quantify the degree of nonclassical dynamics remains. We propose a \emph{nonclassical-impact parameter} (NCI) to characterize the degree of nonclassical effects from an energy perspective, i.e. which fraction of the total electromagnetic energy is attributed to classical electrodynamic terms and which fraction is correspondingly to be assigned to nonclassical degrees of freedom? We show that the NCI relates directly to two fundamental parameters of plasmon resonances: the loss function and the quality factor.
Guided by the NCI, we discuss the nonclassical effects of plasmon waveguiding modes of metallic slab waveguides, and highlight the general features of the nonclassical effects at different microscopic levels by contrasting the numerical results from the semi-classical hydrodynamic Drude model (HDM) and the microscopic random-phase approximation (RPA). The formal relation between the HDM and the RPA is also established for metals by exploring the limit of an infinite work function.
\end{abstract}

\maketitle

\color{black!88!white}

\section{Introduction}

Plasmon resonances, collective oscillations of a free-electron gas against a positive ion background, are subject to significant attentions.\cite{Maier:2007,Gramotnev:2010,Schuller:2010,Brongersma:2015,Baev:2015} Plasmon resonances, including both localized-plasmon resonances and plasmon waveguiding modes, can confine electric fields beyond the optical diffraction limit.\cite{Gramotnev:2010} This leads to numerous applications such as cancer therapy,\cite{Neal:2004} nanophotonics circuits,\cite{Huang2014,Smith:2015} nanolasers,\cite{Noginov:2009,Oulton:2009} and quantum information processing.\cite{Chang:2007} The majority of these developments rely on our insight from classical electrodynamics where the Drude model for the intraband plasmonic response of metals stands as a cornerstone.\cite{Maier:2007,Brongersma:2015} This celebrated approach explains noble-metal plasmon phenomena extremely well for plasmonic structures with characteristic dimensions well above $5{\sim}10\,\rm nm$.\cite{Raza:2015a} However, the maturing nanofabrication allows the realization and exploration of yet smaller feature sizes.\cite{Kern:2012,Savage:2012,Scholl:2013,Raza:2014a,Kern:2015} When approaching the few--nanometer or even sub--nanometer scale, an increasing importance of nonclassical degrees of freedom (i.e. beyond the Drude model) can be anticipated and with increasing weights.\cite{Tame:2013} Examples of such nonclassical degrees of freedom include the kinetics associated with the finite compressibility of the quantum electron gas\cite{Henrik:2004,Raza:2011} and the inhomogeneous microscopic equilibrium distribution of the quantum electron gas in the vicinity of a surface.\cite{Lang:1970,Jin:2015,Yan:2015a} In Maxwell's equations, such aspects modify the local Drude permittivity to a generalized nonlocal form.\cite{Mortensen:2014} With respect to different nonlocal models, such as the semi-classical hydrodynamic Drude model (HDM) and the microscopical random-phase approximation (RPA), the commonly employed Drude model is also referred to as the local--response approximation (LRA).

In recent years, efforts have been directed to theoretically\cite{Abajo:2008,Raza:2011,Nordlander:2012,Esteban:2012,Stella:2013,Mortensen:2014,Teperik:2013,Kulkarni:2015,Townsend:2015} and experimentally\cite{Scholl:2012,Savage:2012,Ciraci:2012b,Raza:2012b,Kern:2012,Mertens:2013,Tan:2014,Raza:2015c,Jung:2015} investigating nonclassical effects of plasmon resonances in metals and with interesting extensions to two--dimensional (2D) plasmonic materials such as graphene.\cite{Thongrattanasiri:2012,Thomas:2014,Wang:2015} In addition to the fundamental interest in quantum--plasmonic phenomena, we note that quantum--electronic control may open new avenues in applications of nanoscale light confinement\cite{Tame:2013} and plasmon-induced hot electrons.\cite{Harutyunyan:2015,Brongersma:2015b} For resonant phenomena, the nonclassical effects usually manifest in broadening and shifting of plasmon resonances.\cite{Scholl:2012,Raza:2012b,Voisin:2000,Feibelman:1982,Apell:1983,Liebsch:1991} In turn, nonclassical effects also smear singularity phenomena predicted by the Drude model, such as perfect imaging,\cite{Larkin:2005} nanofocusing\cite{Marier:2012,Khurgin:2015a} and the field enhancement at the center of a touching dimer\cite{Toscano:2012a,FernandezDominguez:2012} that would otherwise be singular.\cite{Romero:2006} Nonclassical dynamics may also be associated with phenomena without any classical counterparts, such as the mutipole surface-plasmon resonances at the interface of simple metals,\cite{Liebsch:1991} the charge transfer plasmon resonances for dimers with sub-nanometric gaps,\cite{Savage:2012,Scholl:2013,Esteban:2012,Nordlander:2012,Mertens:2013,Kulkarni:2015} and the quantized bulk-plasmon resonances above the plasma frequency.\cite{Ozer:2011}

The theoretical understandings of nonclassical effects of plasmon resonances usually employ the semi-classical HDM,\cite{Toscano:2012a,FernandezDominguez:2012,Toscano:2013,Marier:2012,Bao:2013,Raza:2013w,Mortensen:2014} the RPA,\cite{Ichikawa:2011,Thongrattanasiri:2012,Thomas:2014,Wang:2015} or the time-dependent density-functional theory (DFT).\cite{Liebsch:1991,Nordlander:2012,Teperik:2013,Andersen:2012,Andersen:2013}
Elevating us beyond these different models, it is interesting to ask if we can introduce a measure to quantify the degree of nonclassical contributions to the plasmon dynamics? One immediate candidate at hand is to associate such a parameter with the resonance shift of plasmon modes when compared to the LRA. Obviously, this would work for well--developed resonances existing within the LRA, while it would not help to appreciate quantum plasmon phenomena not holding classical counterparts.
To address a broader variety of nonclassical phenomena, we here propose a \emph{nonclassical-impact parameter} (NCI) to characterize the nonclassical effects from a total energy ($U_{\rm  T}=U_{\rm  C}+U_{\rm  NC}$)  perspective: ${\rm NCI}\equiv  2U_{\rm  NC}/(U_{\rm  C}+U_{\rm  NC})$, where ${\rm NCI}=0$ for classical (C) dynamics, while ${\rm NCI}=1$ for entirely nonclassical (NC) dynamics.
As a key result we show that
\begin{align}
{\rm NCI}=1+\frac{{\rm Im}\left[ \varepsilon_{\rm PR}^{-1}\right]}{Q_{\rm PR}}\label{eq:NCIRPA}
\end{align}
where $Q_{\rm PR}$ is the quality factor of the plasmon resonance while ${\rm -Im}\left[ \varepsilon_{\rm PR}^{-1}\right]$ is the loss function; both established quantities in the fields of plasmonics and electron-energy loss spectroscopy of plasmons.\cite{Abajo:2010}

The remaining part of the paper is organized as follows: In Section II, with the HDM as the starting point, we partition different forms of the energies of plasmon resonances including the contributions from the nonclassical effects. Based on the energy considerations in Section II, we define the NCI to characterize the nonclassical effects beyond the HDM assumption in Section III. In Section IV, we discuss the relation between the NCI and the electron-pressure wave, and demonstrate the fundamental upper limit of the NCI for plasmon waveguiding modes in the HDM.
In Section V, with the NCI, we investigate in detail the nonclassical effects by numerical analysis of plasmon waveguiding modes of metallic slab waveguides. In Section VI, the formal relation between the HDM and the RPA is discussed in the limit of an infinite work function. In Section VII, a summary and conclusions are given. Some details of our derivations are left for Appendices A, B and C.

\section{Nonclassical kinetics in the hydrodynamic Drude model}

We start by introducing the HDM to describe the free-electron gas. In short, the HDM seeks to account for the finite compressibility of the quantum electron gas by incorporating a pressure term in the equation-of-motion of electrons subject to electromagnetic fields. Commonly, Thomas--Fermi theory is used to describe the nonclassical (quantum) freedom of the kinetic energy, which arises from the statistical distribution of the free-electron gas e.g. the Fermi--Dirac statistics in equilibrium.
The internal kinetic energy per particle is proportional to $\max\left\{{\rm E}_{\rm F}, k_{\rm B}T\right\}$, where $\rm E_{\rm F}$ is the Fermi energy, while $k_{\rm B}T$ is the characteristic thermal energy. For metals such as Sodium, Aluminum, Silver and Gold, with the Fermi velocity $v_{\rm F}$ around $10^6\,\rm m/s$ and the effective electron mass close to the electron rest mass, we typically have ${\rm E}_{\rm F} \gg k_{\rm B}T$ at room temperatures,\cite{Henrik:2004} and consequently the nonclassical kinetic energy is governed by $\rm E_{\rm F}$. In the present paper, we focus our attention on the above mentioned metals.
In the HDM, the classical Drude model is accompanied by an additional diffusion-like gradient term\cite{Raza:2011}
\begin{align}
\mathbf J_{\rm e}=\sigma_{\rm D} \mathbf E-\frac{{\rm i}\omega\beta^2}{\omega^2+{\rm i}\omega\gamma}\bm\nabla \rho_{\rm e}.
\label{eq:CurrentHDM}
\end{align}
Here, $\mathbf J_{\rm e}$ and $\rho_{\rm e}$ denote the electric current density and the electric charge density, respectively. Furthermore, $\gamma$ is the damping rate and $\sigma_{\rm D}={\rm i}\varepsilon_0\omega_{\rm p}^2/\left(\omega+{\rm i}\gamma\right)$ is the classical Drude conductivity with $\omega_{\rm p}$ being the plasma frequency. The parameter $\beta$, associated with the nonclassical kinetic energy, has the expression $\beta^2=v_{\rm F}^2\left(3\omega/5+{\rm i}\gamma/3\right)/(\omega+{\rm i}\gamma)$,\cite{Halevi:1995}
and gives the strength of the nonlocal relationship between $\mathbf E$ and $\mathbf J$.
The HDM is usually combined with the assumptions of a uniform equilibrium-electron density and an infinite work function. The two assumptions lead to (1) all material parameters in Eq.~(\ref{eq:CurrentHDM}), i.e., $\sigma_{\rm D}$, $\gamma$, and $\beta$, are set to the bulk values of the corresponding metal; (2) an additional boundary condition is imposed, i.e. the vanishing of the normal component of $\mathbf J_{\rm e}$ at the metal surface.\cite{Raza:2011,Yan:2012}

Before proceeding, we note that the HDM constitutes a lowest-order correction of the LRA to include nonlocal effects due to the finite compressibility of the quantum electron gas. Apart from recent generalizations,\cite{Toscano:2015,Yan:2015b,Li:15} the HDM still neglects several important microscopic effects including the inhomogeneity of the equilibrium-electron density and its related correction to the nonclassical kinetic energy, and the electron spill-out.
Nevertheless, the main motivation here of employing the HDM is to illustrate how the nonclassical degrees of freedom are modifying the dynamics of plasmons. This insight also leads us to our introduction of the $\rm NCI$. At a later stage, the assumptions associated with the HDM will be relaxed.

Now, consider an arbitrary free-electron gas embedded in a dielectric background with a relative permittivity $\varepsilon_{\rm d}$ (possibly varying in space). In the electrostatic limit, the electromagnetic fields associated with plasmon resonances respect\cite{Wang:2006}
\begin{align}
{\int}dv \left[\mathbf E \mathbf D^*+\mathbf E^* \mathbf D\right]=0.\nonumber
\end{align}
Utilizing Eq.~(\ref{eq:CurrentHDM}), the above identity yields
\begin{subequations}
\begin{align}
U_{\rm K}=U_{\rm E}+U_{\rm NC},
\label{eq:Ublance}
\end{align}
with
\begin{align}
&U_{\rm   K}=\frac{1}{4}\int dv_{\rm m} n_0m_{\rm e}|\mathbf v|^2,\\
&U_{\rm  E}=\frac{1}{4}{\int}dv \varepsilon_0\varepsilon_{\rm d}|\mathbf E|^2,\\
&U_{\rm  NC}=\frac{1}{4}{\int}dv_{\rm m} \frac{\beta^2|\rho_{\rm e}|^2}{\varepsilon_0\omega_{\rm p}^2}.
\end{align}
Here, $\int dv_{\rm m}$ is the spatial integration performed over the volume of the free-electron gas, with $n_0$ being the equilibrium-electron density, while $m_{\rm e}$ is the effective electron mass. We note that the same expressions can also be derived from the Poynting theorem.\cite{Jewsbury:1981a} The total energy $U_{\rm  T}$ and the classical energy $U_{\rm  C}$ are
\begin{align}
U_{\rm  T}=U_{\rm  K}+U_{\rm  E}+U_{\rm  NC},\quad  U_{\rm  C}=U_{\rm  K}+U_{\rm  E}.
\label{eq:Utotal}
\end{align}
\end{subequations}

The physical meanings of $ U_{\rm  K,E,NC}$ are indicated by their expressions. In particular, $U_{\rm K}$ is the time average of the classical kinetic energy of the free-electron gas, and $U_{\rm  E}$ relates with the energy of the electric field. $U_{\rm NC}$ is the nonclassical part of the energy. To appreciate the nonclassical nature of $U_{\rm NC}$, we consider a free-electron gas with a density $n=n_0+n_1\cos\left(\omega t\right)$, where $n_0$ and $n_1$ are two constants with $n_1$ being a small perturbation. Assuming that electrons are obeying Fermi--Dirac statics, the system has the nonclassical energy density $3nE_{\rm F}/5$. Accordingly, the time average of the nonclassical kinetic energy density $W_{\rm NC}$ contributed from the dynamic perturbation $n_1$ is
\begin{align}
W_{\rm NC}=\frac{m_{\rm e}v_{\rm F}^2|n_1|^2}{12n_0}+\mathcal{O}(n_1^4/n_0^4).\nonumber
\end{align}
In the limit where $\omega\ll\gamma$ with $\beta^2\simeq v_{\rm F}^2/3$, one finds $U_{\rm NC}={\int}dS_{\rm m} W_{\rm NC}$ by recalling that $\rho_{\rm e}=-e n_1$ and $\omega_{\rm p}^2=n_0e^2/\left(m_{\rm e}\varepsilon_0\right)$. This directly reveals the nonclassical origin of $U_{\rm NC}$. If on the other hand  $\omega\gg\gamma$ with $\beta^2\simeq 3v_{\rm F}^2/5$, then $U_{\rm NC}\simeq 9/5 {\int}dS_{\rm m} W_{\rm NC}$. Here, the prefactor $9/5$ is deviating from $1$ owing to the fact that the Fermi--Dirac statistics no longer applies as $\omega\gg\gamma$.\cite{Slabchov:2010}

\section{Nonclassical-impact parameter}
\label{Sec:NIP}

\subsection{Definition and semi-classical HDM considerations}
The total energy of the plasmon resonance is given by Eq.~(\ref{eq:Ublance}). The Drude model invokes the LRA and is entirely classical, i.e. $U_{\rm NC}=0$. Plasmon resonances are characterized by the harmonic energy transfer between the classical kinetic energy of free electrons and the electric field energy of the whole system, in analogy with optical resonances in dielectrics with the energy transfer between the electric field energy and the magnetic energy.\cite{Khurgin:2015b} Including the nonclassical degrees of freedom, $U_{\rm NC}$ has a final share of the total energy, and this modifies the classical plasmon dynamics accordingly.

Based on the above energy perspective, we here introduce the \emph{nonclassical-impact parameter} (NCI) that we already highlighted in the introduction:
\begin{align}
{\rm NCI}  \equiv \frac{2U_{\rm  NC}}{U_{\rm  C}+U_{\rm  NC}}= 1-\frac{2U_{\rm E}}{U_{\rm  T}}.
\label{eq:NCI}
\end{align}
In the classical Drude model, $\rm NCI=0$ as a direct result of $U_{\rm NC}=0$. Including the nonclassical effects in the HDM, $\rm NCI$ deviates from $0$, and more specifically becomes a positive value. In the limiting case $U_{\rm E}\to 0$ or equivalently $U_{\rm T}\gg U_{\rm E}$, we have ${\rm NCI}\to 1$. In a hydrodynamic description this limit corresponds to the regime where plasmons are fully longitudinal oscillations, i.e. electron-pressure waves as discussed in the next section. The ${\rm NCI}$ is bounded between $0$ and $1$. The larger value of the NCI, the more noticeable nonclassical effects.

\begin{figure}[t!]
\includegraphics[width=0.42\textwidth]{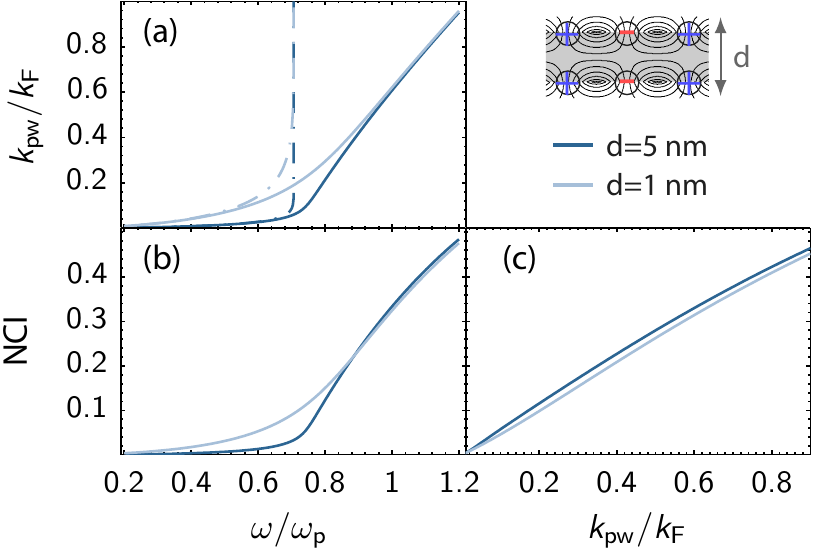}
\caption{The nonclassical effects for the symmetric surface PWMs of a metallic slab waveguide (of width $d=1$\,nm and 5\,nm) imbedded in a free-space background. (a) Dispersion relations within the HDM and the LRA classical Drude model are plotted as the solid lines and the dash-dotted lines, respectively. (b) The NCI illustrated as a function of the plasmon wavenumber. The upper right inset illustrates the mode profile of the symmetric surface PWM. The metal has the material parameters $\omega_{\rm p}=5.9$\,eV and $v_{\rm F}=10^6$\,m/s. }
\label{fig:slab}
\end{figure}

As an example, we consider a metallic slab waveguide imbedded in a free-space background medium. For simplicity we consider a loss-less ($\gamma=0$) free-electron metal with material parameters $\omega_{\rm p}=5.9$\,eV and $v_{\rm F}=10^6$\,m/s (corresponding to Sodium). Fig.~\ref{fig:slab}(a) depicts the dispersions of the symmetric surface plasmon waveguiding modes (PWMs) for waveguide widths of both $d=5$\,nm and $d=1$\,nm, contrasting both the LRA classical Drude model (dash-dotted lines) and the HDM (solid lines). The wavenumber $k_{\rm pw}$ of the waveguide mode is normalized by the Fermi wavenumber $k_{\rm F}$ and likewise, the frequency $\omega$ is conveniently normalized by the plasma frequency $\omega_{\rm p}$. In the Drude model, $k_{\rm pw}$ shows a linear relation with $1/d$,
indicating that $k_{\rm pw}$ can be increased to any value without limitations by decreasing $d$. In the HDM, such linear relation is broken due to nonlocal effects which cut off the large--$k_{\rm pw}$ divergence (see Ref.~\onlinecite{Raza:2013w} for details).
Accordingly, there is a ceiling to $k_{\rm pw}$ (a maximal permitted value of $k_{\rm pw}$), as we will discuss in the next section.
Fig.~\ref{fig:slab}(b) plots the NCI predicted by the HDM as a function of the frequency $\omega$. It is seen that the NCI follows a similar trend as the wavenumber $k_{\rm pw}$ does in Fig.~\ref{fig:slab}(a). In particular, the NCI dramatically increases as $\omega$ exceeds the classical surface-plasmon frequency $\omega_{\rm p}/\sqrt{2}$. Below this frequency the plasmon field is predominantly of a transverse nature in good accordance with the LRA, while it turns gradually more longitudinal beyond this frequency. Fig.~\ref{fig:slab}(c) depicts the NCI as a function of $k_{\rm pw} $. The NCI increases as the $k_{\rm pw}$ increases. This directly illustrates how its nonclassical nature increases as $k_{\rm pw}$ becomes larger. The observation can be directly understood from Eq.~(\ref{eq:CurrentHDM}) where the nonclassical term, i.e., the gradient term, is roughly proportional to $k_{\rm pw}$.

As another example, we consider the NCI for the generic problem of a free-electron metallic sphere imbedded in free space. Using a Mie formulation of the HDM,\cite{Christensen:2014} we find to lowest order in the particle radius $R$ that the NCI for localized plasmon resonances directly manifests the nonlocal blueshifting with
\begin{align}
\text{NCI}&=\frac{\Delta\omega}{\omega_{\rm LRA }}+\mathcal{O}\left(\frac{\beta^2}{\omega_{\rm p}^2R^2}\right)\nonumber\\
          &=\frac{\sqrt{(\ell+1)(2\ell+1)}}{2}\frac{\beta}{\omega_{\rm p}R}+\mathcal{O}\left(\frac{\beta^2}{\omega_{\rm p}^2R^2}\right).\nonumber
\end{align}
Here, $\omega_{\rm LRA}$ is the plasmon resonance frequency in the LRA limit ($\omega_{\rm LRA }=\omega_p/\sqrt{3}$ for the dipole resonance with $\ell=1$), $\Delta\omega$ is the HDM frequency blueshift with respect to $\omega_{\rm LRA }$, and $\ell=1,2,\cdots $ is the angular momentum number of the plasmon resonance. This is a quite intuitive, but also remarkable result which in turn shows how the loss function, the quality factor, and the nonlocal blueshift constitute mutually linked quantities, i.e. $\Delta\omega/\omega_{\rm LRA }\simeq 1+{\rm Im}\left[ \varepsilon_{\rm PR}^{-1}\right]/Q_{\rm PR}$ [see Eq. (\ref{eq:NCIRPA}) or the following subsection for definitions of $\varepsilon_{\rm PR}$ and $Q_{\rm PR}$].

\subsection{Microscopic RPA considerations}
\label{Sec:NIPb}
Above, we used the semi-classical HDM as our starting point for our introduction of the NCI. However, our energy considerations (involving $U_{\rm E}$ and $U_{\rm T}$) go beyond this particular model. Thus, the $\rm NCI$ can also be derived for microscopic quantum models, where more subtle nonclassical effects can be included. Here, we turn to the random-phase approximation as the most common way of calculating the electromagnetic response with a starting point in the single-electron eigenstates of the equilibrium system. In the RPA, the exchange-correlation contributions to the plasmon dynamics are omitted. We note that for our metals of interest with only modest electronic correlations, the induced error is of only quantitative rather than qualitative character.

To discuss the NCI in the RPA, we introduce the spectral representation of the dielectric function operator ${\bepsilona}(\omega)$ with\cite{Andersen:2012,Andersen:2013,Wang:2015}
\begin{align}
\left| \phi_{\rm tot} \right\rangle ={\bepsilona}^{-1}(\omega)\left| \phi_{\rm ext} \right\rangle,\nonumber
\end{align}
where $\left| \phi_{\rm ext} \right\rangle$ and $\left| \phi_{\rm tot} \right\rangle$ represent the states of the external incident and resulting total electric potentials, respectively. Here, $\bepsilonb(\omega)=\mbox{\sffamily\bfseries{I}}-\mbox{\sffamily\bfseries{V}}{\bm\chi_0}(\omega)$, where $\mbox{\sffamily\bfseries{V}}$ is the Coulomb interaction operator defined by $\left\langle\mathbf r\right|\mbox{\sffamily\bfseries{V}}\left| f \right\rangle=-\int d\mathbf r' e^2\left\langle\mathbf r'\right.\left| f \right\rangle/4\pi\varepsilon_0|\mathbf r-\mathbf r'|$, and ${\bm\chi_0}(\omega)$ is defined by $\left\langle\mathbf r\right|{\bm\chi_0}(\omega)\left| f \right\rangle=\int d\mathbf r' \chi_0(\omega,\mathbf r,\mathbf r') \left\langle\mathbf r'\right.\left| f \right\rangle$, where $\chi_0$ is the non-interacting response function constructed from the single-electron orbitals of the underlying equilibrium system.

Plasmon resonances are associated with the poles of ${\bepsilona}(\omega)^{-1}$. In particular, they are the eigenstates of ${\bepsilona}(\omega)$
\begin{align}
{\bepsilona}(\omega)\left| \phi_{n} \right\rangle=\varepsilon_n(\omega)\left| \phi_{n} \right\rangle\nonumber
\end{align}
with $\varepsilon_n(\omega)$ being zero. However, due to the loss, e.g. the electron-momentum relaxation associated with electron-phonon scattering and the electron-hole pair excitations, the pole will be slightly away from the real axis in the complex frequency plane. Since this complicates the evaluation of the complex-valued resonance frequency we here make the simplification to focus on the real part of the frequency (where $\varepsilon_n$ has a complex value). We define plasmon resonances in a loose mathematical sense as the eigenstates with the loss function $-{\rm Im[{\varepsilon_n(\omega)}^{-1}]}$ exhibiting a local spectral peak.\cite{Andersen:2012,Andersen:2013,Wang:2015} Those plasmon resonance (PR) eigenstates and eigenvalues are specifically denoted as $\left| \phi_{\rm PR} \right\rangle$ and $\varepsilon_{\rm PR}$, respectively. The loss function $ {\rm -Im}\left[ \varepsilon_{\rm PR}^{-1}\right]$ is a common concept in electron microscopy of plasmons\cite{Abajo:2010} where it quantifies  the ability of external free carriers (such as a focused electron beam in an electron microscope where the swift electrons act as the sources of $\left| \phi_{\rm ext} \right\rangle$) to couple to plasmons and in this way dissipate energy.
Within the above framework, we find that the NCI can be expressed in a rather elegant form, Eq.~(\ref{eq:NCIRPA}) that is
\begin{align}
{\rm NCI}=1+\frac{{\rm Im}\left[ \varepsilon_{\rm PR}^{-1}\right]}{Q_{\rm PR}},\nonumber
\end{align}
comprising the loss function $-{\rm Im}\left[ \varepsilon_{\rm PR}^{-1}\right]$ and the quality factor $Q_{\rm PR}$, which characterizes the plasmon resonance lifetime. For the details of the derivation we refer to Appendix~B.
Again, we emphasize that in this way our new measure of quantum effects is related directly to long-established key quantities for plasmon resonances; quantities that can in principle be evaluated by various methods and techniques ranging all the way from microscopic theories and \emph{ab-initio} approaches over semi-classical models to even experiments measuring the far-field optical spectra and electron-energy loss spectra of plasmon resonances.

Note that compared to Eq.~(\ref{eq:NCI}) the form of Eq.~(\ref{eq:NCIRPA}) no longer involves  $U_{\rm T}$, for which we do not find a simple expression in the RPA, while quite conveniently the quality factor and the loss function can be evaluated (see Appendix A for details).

Since we have arrived at Eq.~(\ref{eq:NCIRPA}) with more general arguments, the expression is also valid for the classical Drude model and the semi-classical HDM, since both models represent different levels of approximations of the RPA. In particular, the classical Drude model is the local-response approximation of the RPA, while the HDM is a $k^2$-approximation ($k$ being the wavenumber of the electric potential after the Fourier transformation) of the RPA. Thus, Eq.~(\ref{eq:NCIRPA}) offers a generic recipe to quantify the nonclassical effects.

\begin{figure}[t!]
\includegraphics[width=0.42\textwidth]{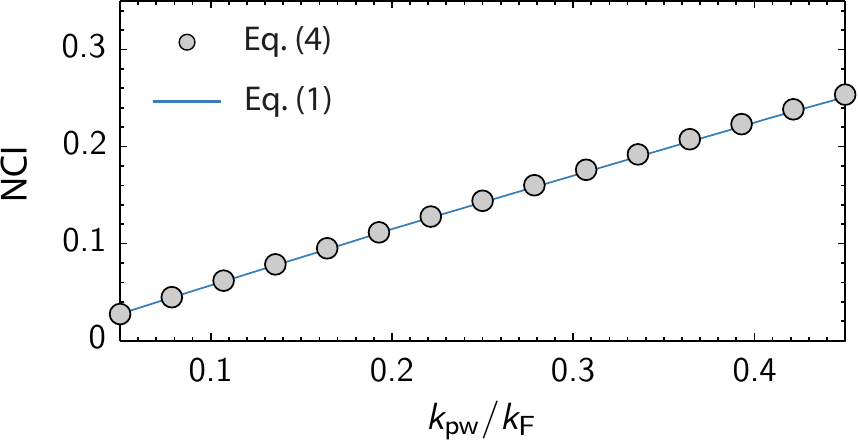}
\caption{Equivalence between Eq.~(\ref{eq:NCIRPA}) [solid line] and Eq.~(\ref{eq:NCI}) [data points] illustrated by HDM calculation of the NCI for a symmetric surface PWM of a metallic 5\,nm slab imbedded in a free-space background. The material parameters of the metal are the same as in Fig.~\ref{fig:slab}. A phenomenological damping rate $\gamma=0.032\omega_{\rm p}$ is added when using Eq.~(\ref{eq:NCIRPA}), while the damping is neglected when evaluating Eq.~(\ref{eq:NCI}). }
\label{fig:Benchmark}
\end{figure}

As an immediate consequence of Eq.~(\ref{eq:NCIRPA}), we have (in agreement with Ref.~\onlinecite{Wang:2006})
\begin{align}
Q_{\rm PR}=-{\rm Im}\left[ \varepsilon_{\rm PR}^{-1}\right]\;\text{ (within the LRA)}
\label{eq:QPDrude}
\end{align}
since the Drude model is classical and $\rm NCI=0$ within the LRA.
As an example, consider the bulk plasmon resonance for a homogenous free-electron gas. In the classical Drude model, the dielectric function operator ${\bepsilona}(\omega)$ is simply the Drude permittivity $\varepsilon_{\rm D}=1-\omega_{\rm p}^2/\omega(\omega+{\rm i}\gamma)$. At the bulk plasmon resonance $\omega=\omega_{\rm p}$, it is evident that $-{\rm Im}\left[ \varepsilon_{\rm PR}^{-1}\right]=\omega_{\rm p}/\gamma$, which also gives the value of the quality factor $Q_{\rm PR}$ according to Eq.~(\ref{eq:QPDrude}). Thus, our considerations are for the LRA in perfect line with the elegant treatment of plasmon resonances in Ref.~\onlinecite{Wang:2006}. Further, to confirm the equivalence between Eqs.~(\ref{eq:NCI}) and (\ref{eq:NCIRPA}), we use the HDM to calculate the NCI of the symmetric surface PWM for a metallic slab waveguide imbedded in a free-space background. The material parameters of the metal are the same as in Fig.~\ref{fig:slab}  and for the width of the waveguide we consider 5\,nm. The results are plotted in Fig.~\ref{fig:Benchmark}, where we observe that the values of the NCI from either Eq.~(\ref{eq:NCI}) or Eq.~(\ref{eq:NCIRPA}) agree perfectly with each other.

\section{Nonclassical-impact parameter and electron-pressure waves}
\label{Sec:Fceil}

In this section, we return to the HDM and discuss the NCI in the context of longitudinal plasmon excitations. In the limiting case with $U_{\rm E}\ll U_{\rm T}$, we have $\text{NCI}\to 1$ as indicated by Eq.~(\ref{eq:NCI}). In this case, plasmon resonances exhibit a harmonic energy transfer mainly between
$U_{\rm K}$ and $U_{\rm NC}$, but with negligible $U_{\rm E}$. Consequently, the electric field term in Eq.~(\ref{eq:CurrentHDM}) can be simply neglected. Then, by applying the gradient operator to Eq.~(\ref{eq:CurrentHDM}) we get
\begin{align}
\left(\nabla^2+k_{\rm P}^2\right)\rho_{\rm e}=0,\quad
k_{\rm P}=\frac{\omega}{\beta}.
\label{eq:PreEq}
\end{align}
The solutions to Eq.~(\ref{eq:PreEq}) are charge-density waves that we shall here refer to as \emph{electron-pressure waves} since the nonclassical gradient term in Eq.~(\ref{eq:CurrentHDM}) is commonly referred to as the pressure term of the free-electron gas.\cite{Slabchov:2010} Thus, as $\text{NCI}\to 1$, the linear-response dynamics of the free-electron gas has the property of an electron-pressure wave.

\begin{figure}[h!]
\includegraphics[width=0.48\textwidth]{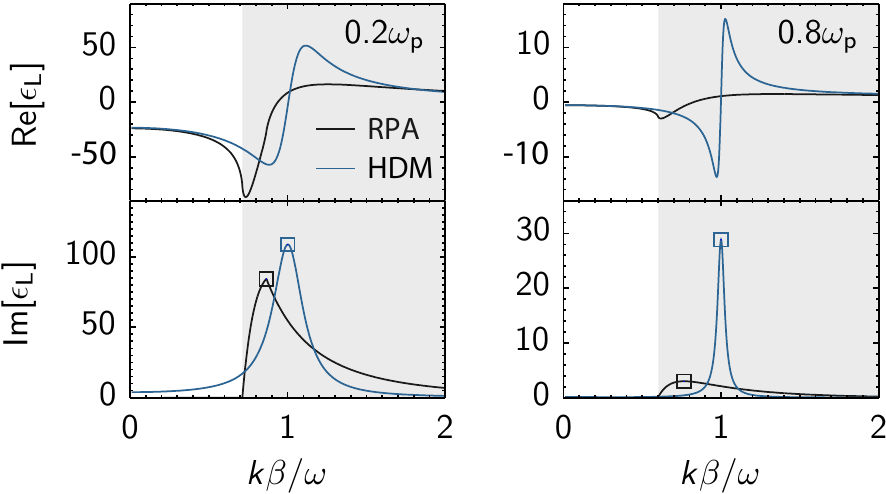}
\caption{
Comparison of the longitudinal permittivities within HDM and LRA at $\omega=0.2\omega_{\rm p}$ (left panels) and  $\omega=0.8\omega_{\rm p}$ (right panels). The material parameters are as in Fig.~\ref{fig:slab}, while using $\gamma=0.032\omega_{\rm p}$ in the HDM. The squares mark the positions of $k_{\rm P}$ while shaded regions represent the RPA intraband electron-hole continuum.}
\label{fig:RH}
\end{figure}

As mentioned above, the HDM is the $k^2$-approximation of the RPA of a homogenous electron gas. Consequently, it is important to see whether the value of $k_{\rm P}$ predicted by the HDM is close to that of the RPA. In this context, we observe that the longitudinal permittivity in the HDM is given by
\begin{align}
\varepsilon_{\rm L}=1-\frac{\omega_{\rm p}^2}{\omega^2-\beta^2 k^2+{\rm i}\omega\gamma}.\nonumber
\end{align}
Clearly, $k_{\rm P}$ is nothing but the resonance pole of $\varepsilon_{\rm L}$ if $\gamma=0$.
Within the RPA this inspires us to extract $k_{\rm P}$ in a similar way by solving for the resonance of $\varepsilon_{\rm L}$. As an example, we considering a metal with properties as in Fig.~\ref{fig:slab}. The left panel of Fig.~\ref{fig:RH} depicts $\varepsilon_{\rm L}$ as a function of $k$ at $\omega=0.2\omega_{\rm p}$ for both the HDM and the RPA. The shaded regions represent the intraband electron-hole (e-h) pair continuum, which contributes to the imaginary part of $\varepsilon_{\rm L}$ in the RPA. In the HDM, a phenomenological damping $\gamma=0.032\omega_{\rm p}$ is added. Clearly, for both the HDM and the RPA, $\varepsilon_{\rm L}$ exhibits the typical resonance features: the abrupt sign change of ${\rm Re }[\varepsilon_{\rm L}]$ accompanied by a peak in ${\rm Im }[\varepsilon_{\rm L}]$. The latter feature is now employed to identify the resonance position (marked by squares), i.e., the value of $k_{\rm P}$.  We see that $k_{\rm P}$ in the RPA is close to that in the HDM with $k_{\rm P}^{\rm RPA}=0.865k_{\rm P}^{\rm HDM}$. The right panel of Fig.~\ref{fig:RH} depicts $\varepsilon_{\rm L}$ as a function of $k$ at $\omega=0.8\omega_{\rm p}$. Comparing with the case of $\omega=0.2\omega_{\rm p}$, we emphasize two observations: (1) $k_{\rm P}$ in the RPA deviates more from that in the HDM with $k_{\rm P}^{\rm RPA}=0.75k_{\rm P}^{\rm HDM}$; (2) the resonance of the electron-pressure wave in the RPA is more damped due to the enhanced e-h pair excitations. In summary, the $k_{\rm P}$ predicted by the HDM is qualitatively accurate and compares well with the RPA, especially in the low-frequency regime.

In Fig.~\ref{fig:Benchmark}, we observe that the NCI increases as the PWM wavenumber $k_{\rm pw}$ grows. Thus, it is reasonable to conjecture that the ceiling to $k_{\rm pw}$ should correspond to the limiting case $\rm NCI=1$, i.e., dynamics characteristic to an electron-pressure wave. This immediately suggests the following relation:
\begin{align}
k_{\rm pw}\le k_{\rm P}.
\label{eq:KspIneq}
\end{align}
For a slab metallic waveguide in a homogenous dielectric background, Eq.~(\ref{eq:KspIneq}) has been demonstrated explicitly by analyzing the dispersion equation.\cite{Yan:2012} It has also been shown that such a ceiling is responsible for the limiting value of the photonic density-of-states of layered hyperbolic metamaterials.\cite{Yan:2012} Here, based on the arguments from the energy perspective, Eq.~(\ref{eq:KspIneq}) is provided without any detailed specifications of either the underlying waveguide geometry or the imbedding medium's dielectric properties. This suggests that Eq.~(\ref{eq:KspIneq}) should be universally valid within the HDM, which is rigorously proven in Appendix~A. To allow $k_{\rm pw}$ to approach $k_{\rm P}$, Fig.~\ref{fig:slab} tells us that one should either make the waveguide's transverse dimensions small or alternatively aim for the high-frequency regime. However, even if the waveguide would hypothetically be as thin as one atomic layer, $k_{\rm pw}$ would still be far away from $k_{\rm P}$, as will be demonstrated in the next section. Additionally, in the high-frequency regime, the enhanced loss due to the e-h pair excitations can reduce the plasmon lifetime greatly (as observed in Fig.~\ref{fig:RH}), which would make experimental observations difficult. In this sense, the fundamental ceiling defined in Eq.~(\ref{eq:KspIneq}) is perhaps not posing a real practical limitation in the presence of realistic damping, but it rather serves as a theoretical concept manifesting the more ultimate limitation associated with the spatial dispersion of electron-pressure waves and their nonclassical kinetic energy.

\section{Numerical Analysis}
\label{Sec:Micro}

The HDM invokes the following approximations and assumptions: (a) the $k^2$-approximation of the RPA; (b) a uniform equilibrium-electron density; (c) an infinite work function. To investigate the nonclassical effects beyond these simplifications, we here employ the microscopic RPA for numerically analyzing the PWMs of metallic slab waveguides.

The RPA automatically lifts the $k^2$-approximation inherent to the HDM. Procedures to further relax the remaining  approximations  and assumptions will depend on the particular choice of the single-electron potential $V_{\rm el}$. Here, two different models for $V_{\rm el}$ are considered. One approach is the infinite work function (IFW) potential well, in which $V_{\rm el}=0$ inside the metal, i.e., the region occupied by the jellium ion background, while $V_{\rm el}\rightarrow \infty$ outside the metal, as illustrated in the left inset of Fig.~\ref{fig:mode}. In this case, the assumption of the uniform electron density is relaxed. This allows for Friedel oscillations in the equilibrium-electron density, while quantum spill-out is suppressed due to the IWF (for an early application of the infinite-barrier idea to plasmons, see Ref.~\onlinecite{Keller:1993}). The other choice involves a single-electron potential $V_{\rm el}$ treated self-consistently within DFT -- a more accurate description,\cite{Kohn:1965} as illustrated in the right inset of Fig.~\ref{fig:mode}. In the DFT, the local-density approximation of the exchange-correlation energy is employed. In this case, the assumptions of the uniform electron density and the infinite work function are both elevated. In the following, our comparison of the two RPA descriptions (the IFW--RPA and the DFT--RPA) and the HDM will exhibit nonclassical effects of plasmon resonances at different microscopic levels.

\subsection{Metallic slab}

\begin{figure}[h!]
\includegraphics[width=0.48\textwidth]{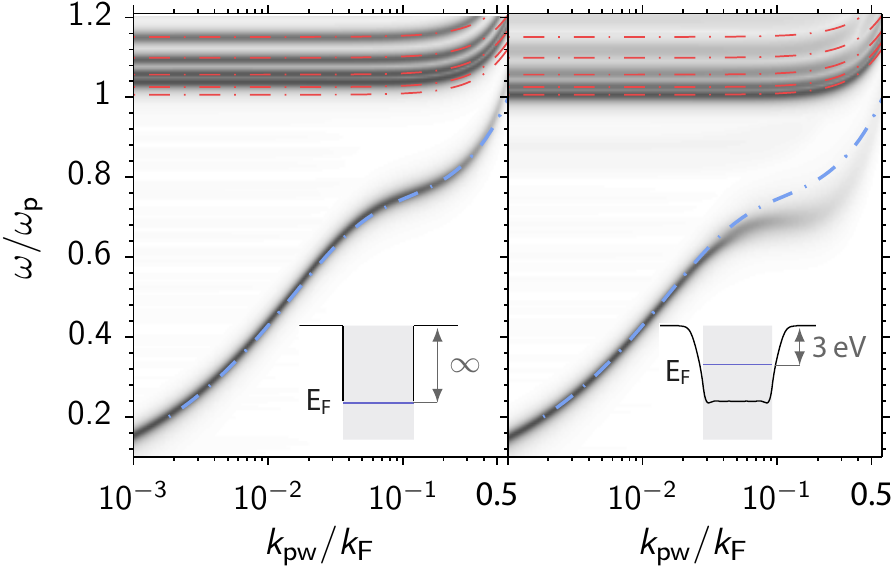}
\caption{Dispersions of the symmetric PWMs for a metallic 5\,nm wide slab waveguide imbedded in free space. The results are computed within the HDM (dash-dotted lines), the IFW--RPA (left) and the DFT--RPA (right). The material parameters of the metal are chosen as in Fig~\ref{fig:slab}.}
\label{fig:mode}
\end{figure}

Figure~\ref{fig:mode} depicts the dispersions of the symmetric PWMs for a metallic slab of width $5\,\rm nm$, contrasting the results from the HDM (dash-dotted lines), the IFW--RPA (left) and the DFT--RPA (right). In the DFT, we employ the jellium approximation for the ion lattices, and self-consistently obtain a work function of 3\,eV (see the inset of the right part of Fig.~\ref{fig:mode}). The material parameters of the metal are as in Fig.~\ref{fig:slab}, while a phenomenological damping rate $\gamma=0.032\,\omega_{\rm p}$ is included. For the IFW--RPA and the DFT--RPA, we illustrate the largest values of the loss function $-{\rm Im[{\varepsilon_n}^{-1}]}$ (see Sec.~\ref{Sec:NIPb}) divided by the frequency, whose local peaks represent the PWMs. The demonstrated PWMs include both the surface PWMs and the bulk PWMs above $\omega_{\rm p}$.

We focus on the surface PWMs (the mode of the low-frequency branch). The results of the HDM and the IFW--RPA agree with each other. This indicates the quantitative agreement between the two models when it comes to characterizing the properties of the surface screening of the free-electron gas. This will be further discussed in Sec.\ref{sec:RHDMRPA}. For the DFT--RPA, the predicted surface PWMs exhibit: (1) redshifting with respect to the HDM and IFW--RPA, (2) more damping as $k_{\rm pw}$ increases. These two observations are attributed to the electron spill-out permitted by a finite work function. In particular, the increased damping relates with the enhanced e-h pair excitations near the jellium boundary (surface scattering) due to the electron spill-out, as recently discussed in Refs.~\onlinecite{Jin:2015,Yan:2015a}.

\begin{figure}[h!]
\includegraphics[width=0.42\textwidth]{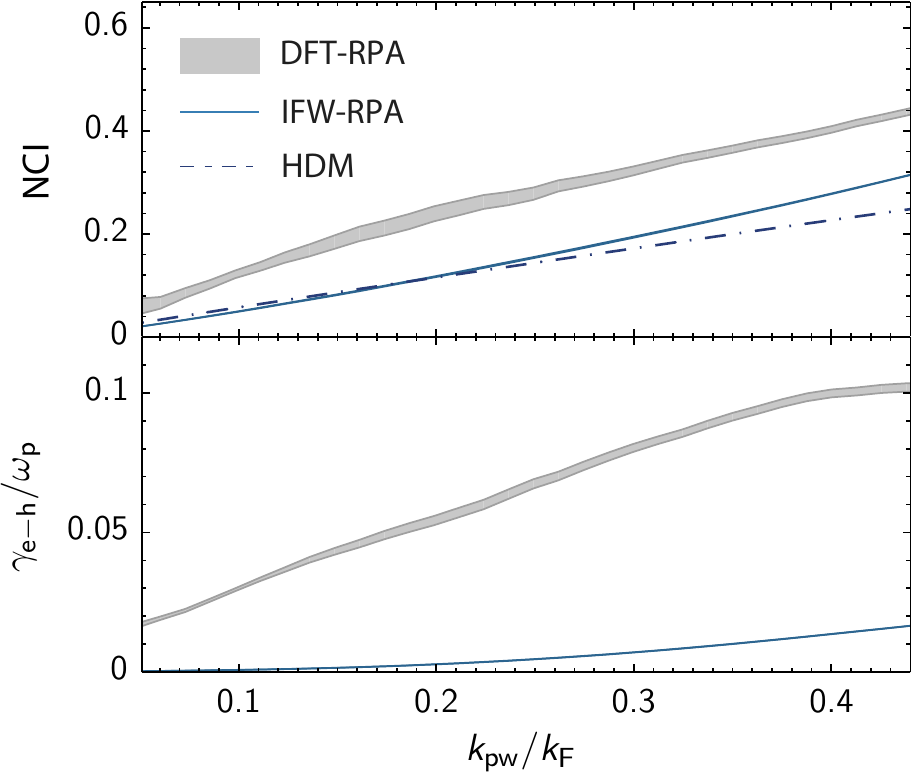}
\caption{(Upper panel) The NCI for the symmetric surface PWMs in Fig.~\ref{fig:mode} contrasting the HDM, the IFW--RPA, and the DFT--RPA results. (Lower panel) $\gamma_{\rm e-h}$ the damping rate of the surface PWMs contributed from the e-h pair excitations contrasting the IFW--RPA and the DFT--RPA.
The DFT--RPA results are illustrated as the stripes with the vertical width representing the numerical uncertainty of determining the quality factor of the PWM.  }
\label{fig:QI1}
\end{figure}

To quantify the nonclassical effects, we next compute the NCI. Fig.~\ref{fig:QI1} depicts the NCI for the surface PWMs considered in Fig.~\ref{fig:mode}, contrasting the HDM, the IFW--RPA, and the DFT--RPA. To compute the NCI in the RPA, the quality factor of the PWMs is needed. While this is indeed feasible, we note that the numerical determination of the quality factor comes with a small numerical uncertainty, especially for the DFT--RPA (see Appendix B for details). To visualize the numerical accuracy, the NCI for the DFT--RPA is accompanied by a shaded region representing the numerical uncertainty. The IFW--RPA gives an NCI close to that of the HDM, which is consistent with the agreement of their mode dispersions in Fig.~\ref{fig:mode}. This suggests that the nonclassical energy of the PWMs in the IFW--RPA is mainly in the form of the hydrodynamic-like nonclassical kinetic energy. The DFT--RPA reveals a higher value of the NCI, which underlines the account for extra nonclassical degrees of freedom. It is reasonable to deduce that such extra degrees of freedom are associated with the finite work function and the associated electron spill-out, which constitute the main difference between the DFT--RPA and the other two models. As an evidence for this, we also plot the damping rate $\gamma_{\rm e-h}$ due to e-h pair excitations in Fig.~\ref{fig:QI1}. It is seen that $\gamma_{\rm e-h}$ is higher within the DFT--RPA than in the IFW--RPA. Thus, for the DFT--RPA, the proportion of energy carried by the e-h pairs is larger, which of course in turn serves to increase the NCI.

\subsection{2D metallic monolayer }

In Section \ref{Sec:Fceil}, we discussed the fundamental ceiling $k_{\rm P}$ to the wavenumber of PWMs as desribed within the HDM. To approach this fundamental ceiling, it is preferable to have metallic waveguides with reduced transverse dimensions. Considering a metallic slab, the thinnest imaginable slab is conceptually that of a two--dimensional (2D) atomic monolayer.\cite{Manjavacas:2014} In this subsection, we will investigate the PWMs of such a monolayer and discuss the practical feasibility of approaching $k_{\rm P}$.

\begin{figure}[h!]
\includegraphics[width=0.48\textwidth]{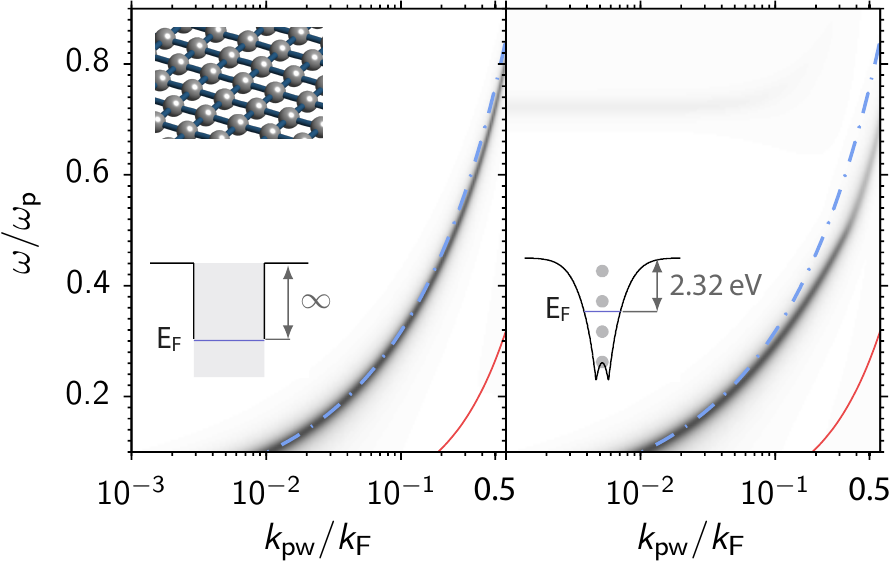}
\caption{Dispersions of the symmetric PWMs for a metallic 2D atomic monolayer. The results are computed within the HDM (dash-dotted lines), the IFW--RPA (left),  and the DFT--RPA (right). The fundamental ceiling for the wavenumber $k_{\rm P}$  is also shown (solid lines). The atomic monolayer is chosen to represent the (100) surface of Sodium, as illustrated in the inset in the left part panel.}
\label{fig:monlayer}
\end{figure}

For the free-electron gas supported by a 2D atomic monolayer, the hydrodynamic equation, Eq.~(\ref{eq:CurrentHDM}), requires a slight modifications owing to the dimensional reduction. In particular, $\mathbf J_e$ and $\rho_e$ in Eq.~(\ref{eq:CurrentHDM}) are the induced electric current density and electric charge density on the 2D plane. Additionally, the Drude conductivity is expressed as $\sigma_{\rm D}={\rm i}n_{\rm 2D}/m_{\rm e}\left(\omega+{\rm i}\gamma\right)$ where $n_{\rm 2D}$ is the equilibrium-electron density of the 2D monolayer. Furthermore, the nonlocal parameter $\beta$ is $\sqrt{3/4}v_{\rm 2DF}$ for $\omega\gg\gamma$, and $\sqrt{1/2}v_{\rm 2DF}$ for $\omega\ll\gamma$, where $v_{\rm 2DF}$ is the Fermi velocity of the 2D free-electron gas.\cite{Fetter:1973,Jackson} Besides the HDM, we also employ the IFW--RPA and the DFT--RPA to characterize the PWMs. In the IFW--RPA, the width of the infinite work function potential well $t_{\rm IFW}$ is chosen by $n_0t_{\rm IFW}=n_{\rm 2D}$. For the DFT--RPA, the ion pseudo-potential proposed by Ashcroft\cite{Ashcroft:1966} is used to describe the ion lattices beyond the jellium approximation. Finally, for numerical convenience we neglect the dependence of the pseudo-potential on the 2D lattice by spatial averaging within the plan while preserving the out-of plane modulation of the potential (see right inset in Fig.~\ref{fig:monlayer}).

The atomic monolayer of the (100) surface of Sodium is taken as a thought example. Such a monolayer would have a square lattice with lattice constant $a_0=4.23 {\AA}$ and accordingly $n_{\rm 2D}=1/a_0^2$. Fig.~\ref{fig:monlayer} depicts the dispersions of the symmetric PWMs, contrasting the HDM (dash-dotted lines), the IFW--RPA (left) and the DFT--RPA (right), and the ceiling wavenumber $k_{\rm P}$ (solid lines). With our eyes guided by the mode dispersion of the HDM,
we observe that the results predicted by the HDM and the IFW--RPA have a good mutual agreement, while the mode dispersion of the DFT--RPA exhibits a redshifting and an enhanced damping as $k_{\rm pw}$ increases. Additionally, we see that the fundamental ceiling wavenumber $k_{\rm P}$ is much larger than $k_{\rm pw}$. This consolidates our previous statement that $k_{\rm P}$  mainly constitutes an ultimate limitation for plasmon focusing beyond the diffraction limit of light.\cite{Yan:2012,Yan:13,Raza:2013w,David:2013}

\begin{figure}[h!]
\includegraphics[width=0.42\textwidth]{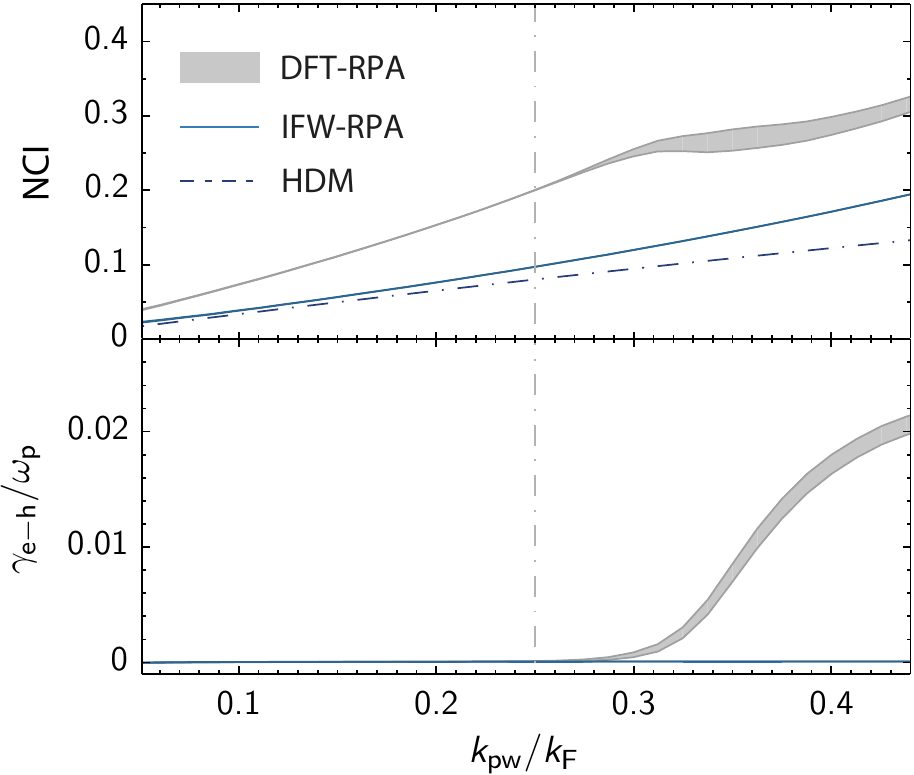}
\caption{The NCI (upper panel) for the symmetric PWMs considered in Fig.~\ref{fig:monlayer}, contrasting the HDM, the IFW--RPA, and the DFT--RPA. The damping rate  $\gamma_{\rm e-h}$  (lower panel) of the surface PWMs contributed from the e-h pair excitations, contrasting the IFW--RPA and the DFT--RPA. For the DFT--RPA results, the shaded region illustrates the numerical uncertainty of determining the quality factor of the PWM. }
\label{fig:QI2}
\end{figure}

Figure~\ref{fig:QI2} depicts the NCI for the PWMs in Fig.~\ref{fig:monlayer}. Similar to our observations in Fig.~\ref{fig:QI1}, the values of the NCI are close to each other for the HDM and the IFW--RPA, while the DFT--RPA predicts a higher value due to the inclusion of additional nonclassical degrees of freedom. In Fig.~\ref{fig:QI2} we also plot $\gamma_{\rm e-h}$, contrasting the IFW--RPA and the DFT--RPA. We see that $\gamma_{\rm e-h}$ for the IFW--RPA is nearly zero and this also holds approximately in the DFT--RPA for $k_{\rm pw}$ smaller than $0.25k_{\rm F}$  (marked by the vertical dash-dotted line). The reason is that the plasmon dynamics of such a monolayer resembles the intraband dynamics of an ideal 2D electron gas as long as the excitation energy is smaller than the minimum energy $E_{\rm int}$ required for e-h interband excitations. For the IFW--RPA the energy should exceed $E_{\rm int}=63.9$\,eV, while for DFT--RPA the energy is much reduced to $E_{\rm int}=1.8$\,eV. Accordingly, when the energy is smaller than $E_{\rm int}$, the e-h pair excitations can only originate from intraband excitations, which is impossible for $k_{\rm pw}<\omega/v_{\rm F2D}$.\cite{Henrik:2004} For the DFT--RPA, the e-h interband excitations are possible when the excitation energy exceeds 1.8\,eV, and
can be further enhanced as the excitation energy approaches and eventually exceeds the work function of 2.32\,eV. This explains our observations discussed earlier in this paper: the $\gamma_{\rm e-h}$ remains low unless $k_{\rm pw}>0.25k_{\rm F}$, in which case the excitation energy favors noticeable e-h interband excitations. Finally, it is interesting to note that the numerical uncertainties of the NCI and $\gamma_{\rm e-h}$ exhibit a correlation with the magnitude of $\gamma_{\rm e-h}$, which is discussed in more detail in Appendix A.

As another example of a 2D material, we briefly discuss doped graphene nanostructures.\cite{Abajo:2014} On the one hand, classical electrodynamics will be unable to resolve atomic-scale details. On the other hand, the underlying graphene lattice can in fact be cut into flakes in two distinct different ways which host slightly different electronic properties: flakes where edge-atoms are configured in either an armchair or zigzag fashion. The latter is giving rise to localized electronic edge states not supported by the former. These zigzag edge-states host additional optical transitions not found for armchair structures.\cite{Thomas:2014} This is indeed also seen for the plasmon spectra of ribbons\cite{Thongrattanasiri:2012} and triangularly shaped flakes,\cite{Wang:2015} where zigzag structures exhibit additional quantum dynamics not exhibited by the armchair counterparts.\cite{Thomas:2014} In fact, the different magnitude in loss functions reported in Ref.~\onlinecite{Wang:2015} are in agreement with our anticipation that the zigzag structures exhibit a larger NCI than their armchair counterparts.

\section{Discussion}
\label{sec:RHDMRPA}
The numerical results of the PWMs for a metallic slab show an excellent agreement between the HDM and the IFW--RPA, see Figs.~\ref{fig:mode} and \ref{fig:monlayer}. Is this just by pure accidence or is it a manifestation of two more closely connected approaches? Here, we argue that the latter is indeed the case. We note that a formal relation between the HDM and the RPA is already established in Section II B of Ref.~\onlinecite{Efguiluz:1979}, while we in Appendix~\ref{app:RHDMRPA} revisit the problem to offer a more explicit connection between the HDM and the IFW--RPA.

It should be emphasized that the (semi-)infinite electron gas obviously does not support any discretization of electron orbitals, i.e., quantum-size effects associated with the particle-in-a-box picture.\cite{Abajo:2012} For an arbitrary confined electron gas occupying a finite region of space, quantum size effects would be captured in the RPA of the bounded electron gas, but not in the corresponding HDM. This can lead to an additional discrepancy between the HDM and RPA for the bounded electron gas. However, only minor differences are to be anticipated if quantum confinement is limited, i.e. for dimensions of the metal volume significantly exceeding the Fermi wavelength.

For noble metals, the Fermi wavelength is sub-nanometric ($\lambda_F\sim 0.5$\,nm) and in few-nanometer sized metal particles, thermal smearing would typically exceed the level spacing of the particle-in-a-box picture. For weakly doped semiconductors and low-dimensional materials, the Fermi wavelength can be much longer and quantization effects can indeed appear (quantized conductance is a famous consequence of this\cite{vanWees:1988,Tombros:2011}). Additionally, in the extreme case like a 2D monolayer where strong out-of-plane quantum confinement may serve to only have electrons occupying a single 2D sub-band, the HDM for 2D in-plane bounded electron gas again serves the $k^2$-approximation of the RPA for the corresponding bounded electron problem. From this perspective, the observed numerical agreement between the HDM and the IFW--RPA in Figs.~\ref{fig:mode} and \ref{fig:monlayer} is to be expected. For doped graphene nanostructures (without edge-state contributions associated with zigzag edge termination) we have also reported similar good agreement between HDM and RPA.\cite{Thomas:2014}

\section{Conclusions}
In conclusion, we have proposed a \emph{nonclassical-impact parameter} (NCI) to quantify the nonclassical effects of plasmon resonances from an energy perspective. Importantly, we have provided a general expression for the NCI which links up to quantities commonly employed within classical electrodynamics in the discussion of plasmon resonances: the loss function and the quality factor. We have discussed the relation between the limiting value of the NCI and the electron-pressure wave in the hydrodynamic Drude model (HDM) and explored the ultimate ceiling to the confinement of the plasmon waveguiding modes. Guided by the NCI, nonclassical effects have been explored numerically for plasmon waveguiding modes, contrasting the HDM and the microscopic random-phase approximation (RPA). Finally, we have detailed the formal relation between the HDM and the RPA for a metal slab represented as an infinite work function potential well.

\begin{acknowledgments}
We acknowledge Thomas Christensen for catalyzing our conceptual development of the nonclassical-impact parameter and thank Martijn Wubs for insightful comments on our manuscript. The Center for Nanostructured Graphene is sponsored by the Danish National Research Foundation, Project DNRF58. The work was also supported by the Danish Council for Independent Research--Natural Sciences, Project 1323-00087. W.Y. is financially supported by the Lundbeck Foundation, grant no.~70802.
\end{acknowledgments}

\appendix

\section{Derivation of Eq.~(\ref{eq:NCIRPA})}

For plasmon resonances, the total energy $U_{\rm T}$ and the power dissipation $P_{\rm L}$ are related by the quality factor $Q_{\rm PR}$ through the relation
\begin{align}
U_{\rm T}=\frac{P_{\rm L}Q_{\rm PR}}{\omega_{\rm PR} }.
\label{eq:UTn}
\end{align}
Thus, we already now see how the quality factor enters Eq.~(\ref{eq:NCI}). Next, as mentioned in Section \ref{Sec:Micro}, plasmon resonances relate with the eigenstate of the dielectric function operator $\bepsilona$, which in its spectral representation can be expressed as\cite{Andersen:2012,Andersen:2013}
\begin{align}
\bepsilona(\omega)=\sum_n \varepsilon_n(\omega)\left|\phi_n(\omega)\right\rangle\left\langle\tilde\rho_n(\omega) \right|.\nonumber
\end{align}
Here $\left|\phi_n(\omega)\right\rangle$ represents the eigenstate of $\bepsilona$. The $\left |\phi_n(\omega)\right\rangle$ and $\left |\rho_n(\omega)\right\rangle$ satisfy the potential--density relation
\begin{align}
\bm\nabla\cdot\varepsilon_{\rm d}(\mathbf r)\bm\nabla \left\langle \mathbf r |\phi_n(\omega)\right\rangle=-\frac{\left\langle \tilde\rho_n(\omega) | \mathbf r \right\rangle}{\varepsilon_0}.\nonumber
\end{align}
Among $\left |\phi_n(\omega)\right\rangle$, the electric potential state of the plasmon resonance is denoted as $\left |\phi_{\rm PR}(\omega_{\rm PR})\right\rangle$. The eigenvalue is $\left |\rho_{\rm PR}\right\rangle$ with ${\rm Im}[\varepsilon_{\rm PR}^{-1}]$ exhibiting a peak at $\omega_{\rm PR}$. The related electron charge density state $\left |\rho_{\rm PR}\right\rangle$ at $\omega_{\rm PR}$ is defined by
\begin{align}
\left|\rho_{\rm PR}\right\rangle=\left \langle \tilde\rho_{\rm PR}\right|.\nonumber
\end{align}
Next, consider an arbitrary external electric potential state  $\left|\phi_{\rm ext} \right\rangle$ at the resonant frequency $\omega_{\rm PR}$ incident on the plasmonic structure. The induced potential state of plasmon resonances, denoted as $\left| \phi_{\rm PR}^{\rm i} \right\rangle$, has the expression
\begin{align}
\left| \phi_{\rm PR}^{\rm i} \right\rangle&=\varepsilon_{\rm PR}^{-1}\frac{\left\langle \tilde \rho_{\rm PR} |\phi_{\rm ext} \right\rangle}{\left\langle \tilde \rho_{\rm PR} |\phi_{\rm PR} \right\rangle}\left| \phi_{\rm PR} \right\rangle\nonumber.
\end{align}
The dissipation energy contributed from $\left| \phi_{\rm PR}^{\rm i} \right\rangle$ is
\begin{align}
P_{\rm L}=-\frac{1}{2}{\rm Im}\left[\varepsilon_{\rm PR}^{-1}\omega_{\rm PR}\frac{\left\langle \tilde \rho_{\rm PR} |\phi_{\rm ext} \right\rangle}{\left\langle \tilde \rho_{\rm PR} |\phi_{\rm PR} \right\rangle}\left\langle \phi_{\rm ext}|\rho_{\rm PR}  \right\rangle
\right],
\end{align}
and the electric field energy associated with $\left| \phi_{\rm PR}^{\rm i} \right\rangle$ is
\begin{align}
U_{\rm E}=\frac{1}{4}|\varepsilon_{\rm PR}^{-1}|^2 \left|\frac{\left\langle \tilde \rho_{\rm PR} |\phi_{\rm ext} \right\rangle}{\left\langle \tilde \rho_{\rm PR} |\phi_{\rm PR} \right\rangle}\right|^2\left\langle  \phi_{\rm PR} |\rho_{\rm PR} \right\rangle.
\end{align}
With Eq.~(\ref{eq:UTn}), expressions for $P_{\rm L}$ and $U_{\rm T}$ can be derived and Eq.~(\ref{eq:NCI}) is eventually rewritten as
\begin{align}
{\rm NCI}=1-\frac{1}{Q_{\rm PR}}{\rm Im}\left[ \left(\varepsilon_{\rm PR}^{-1}\right)^*\frac{\left\langle  \phi_{\rm ext} |\rho_{\rm PR} \right\rangle
\left\langle  \phi_{\rm PR} |\tilde\rho_{\rm PR} \right\rangle}{\left\langle\phi_{\rm ext} |\tilde\rho_{\rm PR}\right\rangle\left\langle  \phi_{\rm PR} |\rho_{\rm PR} \right\rangle}\right].
\label{eq:NCIa1}
\end{align}
To excite plasmon resonances, the most efficient way is to choose $\left|\phi_{\rm ext}\right\rangle\propto\left|\phi_{\rm PR}\right\rangle$. In this way we finally arrive at  Eq.~(\ref{eq:NCIRPA}).

We note that the NCI is an intrinsic property of plasmon resonances, and should be independent of the external potential.  However, the NCI in Eq.~(\ref{eq:NCIa1}) shows a dependence on the external potential, and Eq.~(\ref{eq:NCIRPA}) is derived with a special choice of $\left|\phi_{\rm ext}\right\rangle$. This is because $\left|\phi_{\rm PR}\right\rangle$ is only a close approximation to plasmon resonances with losses. Evidently, if the loss is zero, we have $\left|\rho_{\rm PR}\right\rangle=\left|\tilde\rho_{\rm PR}\right\rangle$, and the dependence on the external potential of the NCI is removed. The final expression of the NCI is again Eq.~(\ref{eq:NCIRPA}).

To compute the NCI, $Q_{\rm PR}$ and ${\rm Im}\left[ \varepsilon_{\rm PR}^{-1}\right]$ are needed. ${\rm Im}\left[ \varepsilon_{\rm PR}^{-1}\right]$ can be directly computed by solving the eigenvalues of the operator $\bepsilona$. For $Q_{\rm PR}$, it can be expressed as
\begin{align}
Q_{\rm PR}=\frac{\omega_{\rm PR}}{\gamma_{\rm PR}},
\end{align}
with $\text{$\gamma_{\rm PR}$ defined as the resonance width of $-{\rm Im}[\varepsilon_{\rm PR}^{-1}]$ }.$
To determine $\gamma_{\rm PR}$, we use a Lorentzian function to fit the spectrum of  ${\rm Im}\left[ \varepsilon_{n}^{-1}(\omega)\right]$, and the width of the fitted Lorentzian function is $\gamma_{\rm PR}$. From the numerical observations, it is found that such a scheme to determine $\gamma_{\rm PR}$ is appropriate for the IFW--RPA with the fitting error below $1\%$. However, for the DFT--RPA, the fitting error for parameters in Figs.~\ref{fig:QI1} and \ref{fig:QI2} is larger with the highest value approaching approximately $4\%$. The fitting error gives the numerical uncertainty of $\gamma_{\rm PR}$, which is visualized by the shaded regions in Figs.~\ref{fig:QI1} and \ref{fig:QI2}. Further, we deduce that the larger fitting error in the DFT--RPA relates with the stronger e-h excitations, which distort the resonance spectrum from the Lorentzian shape. The deduction is indirectly evidenced by the observations in  Fig.~\ref{fig:QI2} that the numerical uncertainty increases with $\gamma_{\rm e-h}$.

\section{Proof of Eq.~(\ref{eq:KspIneq})}

To prove Eq.~(\ref{eq:KspIneq}), we first rewrite Eq.~(\ref{eq:CurrentHDM}) in terms of the polarization field $\mathbf P$ defined by $\mathbf P={\rm i}\varepsilon_0\mathbf J_{\rm e}/\omega$ and the electric field $\mathbf E$:
\begin{align}
\bm P+\frac{\beta^2}{\omega^2}\bm\nabla\bm\nabla\cdot\bm P=-\frac{\omega_{\rm p}^2}{\omega^2}\bm E.
\label{eq:HDMP}
\end{align}
Here, we have without loss of generality suppressed the damping $\gamma$. Next, by applying the operation ${\int} ds_{\rm m}\,{\rm P}_z^*\hat {\mathbf z}\cdot$ to both sides of Eq.~ (\ref{eq:HDMP}) we arrive at
\begin{align}
k_{\rm pw}^2=k_{\rm P}^2+\frac{\displaystyle\int ds_{\rm m} \omega_{\rm p}^2{\rm E}_z{\rm P}_z^*/\beta^2-{\rm i}k_{\rm pw}\mathbf P_{  ||}\cdot\boldsymbol\nabla_{  ||}
{\rm P}_z^*}{{\displaystyle\int ds_{\rm m} |{\rm P}_z|^2}}.
\label{eq:KspHDM}
\end{align}
Here, the cross-sectional plane of the waveguide is chosen to be the $x$-$y$ plane and subscripts are used to indicate vector components parallel to the cross-sectional plane. To derive Eq.~(\ref{eq:KspHDM}), we employed the additional boundary condition that the normal component of $\mathbf P$ is vanishing at the boundary. Next, we focus on the integrand (in the following denoted by $p$) in the numerator on the right-hand side of Eq.~(\ref{eq:KspHDM}):
\begin{align}
p=\frac{\omega_{\rm p}^2}{\beta^2}p_1+p_2,
\end{align}
where
\begin{subequations}
\begin{align}
p_1&={\int}ds_{\rm m}\, {\rm E}_z{\rm P}_z^*,\nonumber\\
p_2&={\int}ds_{\rm m}\, -{\rm i}k_{\rm pw}\mathbf P_{  ||}\cdot\boldsymbol\nabla_{  ||}
{\rm P}_z^*.\nonumber
\end{align}
 \end{subequations}

For $p_1$, we may write the electrical field as ${\rm E}_z=-{\rm i}k_{\rm pw}\phi$, where $\phi$ represents the electric potential, and for the polarization we have ${\rm P}_z={\rm i}k_{\rm pw}\phi \omega_{\rm p}^2/\omega^2-{\rm i}k_{\rm pw} \nabla^2\phi\beta^2/\omega^2$ from Eq.~(\ref{eq:HDMP}). In this way we have
\begin{align}
p_1&=-\displaystyle\int ds_{\rm m}\left(\frac{k_{\rm pw}^4\beta^2+k_{\rm pw}^2\omega_{\rm p}^2}{\omega^2}|\phi|^2-\frac{k_{\rm pw}^2\beta^2}{\omega^2}\phi\nabla_{   ||}^2\phi^*\right).\nonumber
\end{align}
Using the identity $\phi\nabla_{   ||}^2\phi^*=\boldsymbol\nabla_{   ||}\cdot\left(\phi\boldsymbol\nabla_{   ||}\phi^*\right)-|\boldsymbol\nabla_{   ||}\phi|^2$, the integral $\int ds_{\rm m} \phi\nabla_{   ||}^2\phi^* $ is found to be
\begin{align}
-\int ds_{\rm m}\,|\bm \nabla_{   ||}\phi|^2{+}\oint dl_{\rm m} \phi\boldsymbol_{   ||}\nabla\phi^*\cdot \hat {\mathbf n}.\nonumber
\end{align}
Here, $\oint dl_{\rm m}$ is the integration along the boundary of the metal side, and $\hat{\mathbf n}$ is the surface normal pointing outward the metal. Employing the boundary conditions that $\phi_{\rm m}=\phi_{\rm d}$ and $\boldsymbol\nabla_{   ||}\phi_{\rm m}\cdot\hat{\mathbf n}=\varepsilon_{\rm d}\boldsymbol\nabla_{   ||}\phi_{\rm d}\cdot\hat{\mathbf n}$, where the subscripts "m" and "d" indicate the quantities at the boundary belonging to the metal and the imbedding dielectric medium, respectively, the integration $\oint dl_{\rm m} \phi\boldsymbol_{   ||}\nabla\phi^*\cdot \hat {\mathbf n}$ can be rewritten as
\begin{align}
\int ds_{\rm d} \varepsilon_{\rm d} \left(|\bm \nabla_{   ||}\phi|^2+\varepsilon_{\rm d}k_{\rm pw}^2|\phi|^2\right)\nonumber.
\end{align}
In summary, we have
\begin{align}
p_1=&-\int ds_{\rm m}\left(\frac{k_{\rm pw}^4\beta^2+k_{\rm pw}^2\omega_{\rm p}^2}{\omega^2}|\phi|^2+\frac{k_{\rm pw}^2\beta^2}{\omega^2}|\bm\nabla_{   ||}\phi|^2\right)\nonumber\\
&-\int ds_{\rm m}\left(\frac{\varepsilon_{\rm d}k_{\rm pw}^4\beta^2}{\omega^2}|\phi|^2+\frac{\varepsilon_{\rm d}k_{\rm pw}^2\beta^2}{\omega^2}|\bm\nabla_{   ||}\phi|^2\right).\nonumber
\end{align}
For the dielectric background with $\varepsilon_{\rm d}>0$, it is clear that $p_1<0$.

 Next, we turn to $p_2$. The potential $\phi$ in the metal can be decomposed into the transverse component $\phi^{\rm T}$ and the longitudinal component $\phi^{\rm L}$, respectively. These potentials satisfy $\nabla^2\phi^{\rm T}=0$ and $\nabla^2\phi^{\rm L}+k_{\rm L}^2\phi^{\rm L}=0$, where $k_{\rm L}=\sqrt{\omega^2-\omega_{\rm p}^2}/\beta$. In terms of $\phi^{\rm T}$ and $\phi^{\rm L}$, $\mathbf P$ is expressed as
\begin{align}
\mathbf P=(\bm\nabla_{  ||}+{\rm i}k_{\rm pw}\hat{\mathbf z})\left[\frac{\omega_{\rm p}^2}{\omega^2}\phi^{\rm T}+\phi^{\rm L}\right].\nonumber
\end{align}
The above expression directly leads to
\begin{align}
p_2&=-{\int}ds_{\rm m}\,k_{\rm pw}^2\left|\frac{\omega_{\rm p}^2}{\omega^2}\boldsymbol\nabla_{  ||}\phi^{\rm T}+\boldsymbol\nabla_{  ||}\phi^{\rm L}\right|^2,\nonumber
\end{align}
which implies that $p_2<0$. With Eq.~(\ref{eq:KspHDM}) and the inequalities $p_1<0$ and $p_2<0$, we now finally arrive at Eq.~(\ref{eq:KspIneq}).

\section{HDM versus IFW--RPA}
\label{app:RHDMRPA}

First, considering an extended electron gas with a uniform equilibrium-electron density. In this case, it is known that the HDM is the $k^2$-approximation of the RPA. We represent this symbolically as
\begin{align}
\text{RPA}_{\rm o}\quad\scalebox{1.4}{$ \xrightarrow{k^2\rm-approx.}$}\quad\text{HDM}_{\rm o}.
\label{eq:RPAHDM0}
\end{align}
with the subscript reminding us of the underlying equilibrium assumption of an infinite homogenous electron gas.

Next, consider the bounded electron gas, where the spatial extension is now finite. In this case, the HDM assumes a uniform equilibrium-electron density while the act of an infinite work function is incorporated in the equations through an additional boundary condition that accounts for the vanishing electron flux through the surface of the metal. For clarity, we symbolically denote this model as "HDM$_{\rm b}$" in order to clearly distinguish it from the case of the infinite electron gas. To establish the relation between the HDM$_{\rm b}$ and the RPA, we turn to the HDM$_{\rm b}$ in an integral form
\begin{align}
\rho_{\rm e}(\mathbf r)=\int d\mathbf r' \scalebox{1.1}{$\chi$}_{\rm HDM_b}(\omega,\mathbf r,\mathbf r') \left[e^2\phi(\mathbf r')\right]\nonumber
\end{align}
resembling the form within RPA. For an arbitrary electron gas, we do not find a general way to derive the expression of $\scalebox{1.1}{$\chi$}_{\rm HDM_b}$. To make progresses, we put the generality aside, and focus on the specific case of a semi-infinite electron gas with a planar boundary. The new light shed by this specific example will then guide us to a more general conclusion. The semi-infinite electron gas is located at $x\leq0$. With Eq.~(\ref{eq:CurrentHDM}) and the additional boundary condition, $\scalebox{1.1}{$\chi$}_{\rm HDM_b}$ is then derived:
\begin{align}
\scalebox{1.1}{$\chi$}_{\rm HDM_b}(\omega,\mathbf r,\mathbf r')=\scalebox{1.1}{$\chi$}_{\rm HDM_o}(\omega,\mathbf r,\mathbf r')+\scalebox{1.1}{$\chi$}_{\rm HDM_o}(\omega,\mathbf r,{\mathcal{P}}_x\mathbf r').
\label{eq:xHDMb}
\end{align}
Here, $\scalebox{1.1}{$\chi$}_{\rm HDM_o}$ is the response function of the HDM for the infinite homogenous electron gas, and ${\mathcal{P}}_x$ is the parity operator on the $x$ coordinate defined by ${\mathcal{P}}_x\mathbf r=\left\{-x,y,z\right\}$. The first response function has the expression
\begin{align}
\scalebox{1.1}{$\chi$}_{\rm HDM_o}(\omega,\mathbf r,\mathbf r')=&-\frac{n_0}{m_e\beta^2}\delta(\mathbf r-\mathbf r')\nonumber\\
&-\frac{n_0k_{\rm P}^2}{4\pi m_e\beta^2|\mathbf r-\mathbf r'|}e^{{\rm i}k_{\rm P}|\mathbf r-\mathbf r'|},\nonumber
\end{align}
where $k_{\rm P}$ is the wavenumber of the electron-pressure wave defined in Eq.~(\ref{eq:PreEq}). Eqs.~(\ref{eq:RPAHDM0}) and (\ref{eq:xHDMb}) together suggest that the HDM$_{\rm b}$ can be considered as the $k^2$-approximation of a specific RPA denoted as RPA$_{\rm b}$, with the response function $\scalebox{1.1}{$\chi$}_{\rm RPA_b}$ being
\begin{align}
\scalebox{1.1}{$\chi$}_{\rm RPA_b}(\omega,\mathbf r,\mathbf r')=\scalebox{1.1}{$\chi$}_{\rm RPA_o}(\omega,\mathbf r,\mathbf r')+\scalebox{1.1}{$\chi$}_{\rm RPA_o}(\omega,\mathbf r,{\mathcal{P}}_x\mathbf r').
\label{eq:RPAb}
\end{align}
Here, $\scalebox{1.1}{$\chi$}_{\rm RPA_o}$ is the response function of the RPA for the infinite homogenous electron gas. With the expression of $\scalebox{1.1}{$\chi$}_{\rm RPA_o}$, the $\scalebox{1.1}{$\chi$}_{\rm RPA_b}$ can be written as
\begin{align}
\scalebox{1.1}{$\chi$}_{\rm RPA_b}=\frac{\scalebox{1.1}{$\chi$}_{\rm IFW}}{2}+\frac{\scalebox{1.1}{$\chi$}_{\rm ZMB}}{2}
\end{align}
width $\scalebox{1.1}{$\chi$}_{\rm IFW}$ being the response function for the IFW--RPA, whose physical meaning is presented in Sec.\ref{Sec:Micro}. In the second term, $\scalebox{1.1}{$\chi$}_{\rm ZMB}$ is the response function for the bounded electron gas satisfying the boundary condition that the normal component of the derivative of the electron wave function is vanishing at the surface. This boundary condition is mathematically well defined, provided that the electron mass approaches zero at the boundary, i.e., a \emph{zero-mass boundary} (ZMB). The relation between the HDM$_{\rm b}$ and the RPA$_{\rm b}$ for the semi-infinite electron gas can now be symbolically summarized as
\begin{align}
\frac{\text{IFW--RPA}}{2}+\frac{\text{ZMB--RPA}}{2}\quad\scalebox{1.4}{$\xrightarrow{k^2\rm-approx.}$}\quad\text{HDM}_{\rm b}.
\label{eq:RPAHDM}
\end{align}

Below, we provide an intuitive shortcut to better appreciate Eq.~(\ref{eq:RPAHDM}). First, in the IFW--RPA and the ZMB--RPA, the electron wave functions $\psi$ are determined by the boundary conditions $\psi_{\rm IFW}(x=0)=0$ and $\partial \psi_{\rm ZMB} (x=0)/\partial x=0$, respectively. Thus, we have $\psi_{\rm IFW}\propto \sin(k_x x)e^{{\rm i}\mathbf k_{||}\cdot\mathbf r_{||}}$ and $\psi_{\rm ZMB}\propto \cos(k_x x)e^{{\rm i}\mathbf k_{||}\cdot\mathbf r_{||}}$, where the subscripts indicate the parallel components of the vectors in the $y$-$z$ plane. The wave functions ensure that the normal component of the electron current vanishes at the boundary, which is consistent with the additional boundary condition in the HDM$_{\rm b}$. Additionally, a uniform equilibrium-electron density is assumed within the HDM$_{\rm b}$. This feature is respected by the average of the equilibrium-electron density of the IFW--RPA and the ZMB--RPA, i.e., $n_0(x)\propto |\psi_{\rm IFW}|^2+|\psi_{\rm ZMW}|^2\propto \cos^2(k_x x)+\sin^2(k_x x)\propto \rm constant$. Thus, important features of the HDM$_{\rm b}$ are all covered by the average of the IFW--RPA and the ZMB--RPA, which highlights the validity of Eq.~(\ref{eq:RPAHDM}).

\bibliographystyle{apsrev4-1}
%

\end{document}